\documentclass{aastex631}
\usepackage{amsmath}

\newcommand{\DataDate}{2023 Apr 5}
\newcommand{\NumHJs}{137}

\newcommand{\YeedPdE}{$86.7\,{\rm \mu s\ orbit^{-1}}$}

\revised{2023 Jul 23}
\accepted{2023 Aug 7}

\begin{document}

\title{Metrics for Optimizing Searches for Tidally Decaying Exoplanets}

\author[0000-0002-9495-9700]{Brian Jackson}
\affiliation{Department of Physics, Boise State University\\ 
1910 University Drive, \\ 
Boise ID 83725-1570 USA}

\author[0000-0002-9131-5969]{Elisabeth R. Adams}
\author[0000-0003-3716-3455]{Jeffrey P. Morgenthaler}
\affiliation{Planetary Science Institute\\ 
1700 E. Ft. Lowell, Suite 106, \\
Tucson, AZ 85719, USA}

\begin{abstract}
Tidal interactions between short-period exoplanets and their host stars drive orbital decay and have likely led to engulfment of planets by their stars. Precise transit timing surveys, with baselines now spanning decades for some planets, are directly detecting orbital decay for a handful of planets, with corroboration for planetary engulfment coming from independent lines of evidence. More than that, recent observations have perhaps even caught the moment of engulfment for one unfortunate planet. These portentous signs bolster prospects for ongoing surveys, but optimizing such a survey requires considering the astrophysical parameters that give rise to robust timing constraints and large tidal decay rates, as well as how best to schedule observations conducted over many years. The large number of possible targets means it is not feasible to continually observe all planets that might exhibit detectable tidal decay. In this study, we explore astrophysical and observational properties for a short-period exoplanet system that can maximize the likelihood for observing tidally driven transit-timing variations. We consider several fiducial observational strategies and real exoplanet systems reported to exhibit decay. We show that moderately frequent (a few transits per year) observations may suffice to detect tidal decay within just a few years. Tidally driven timing variations take time to grow to detectable levels, and so we estimate how long that growth takes as a function of timing uncertainties and tidal decay rate and provide thresholds for deciding that tidal decay has been detected.
\end{abstract}


\section{Introduction} \label{sec:Introduction}

From the discovery of the first exoplanet orbiting a Sun-like star \citep{1995Natur.378..355M}, orbital decay powered by tidal interactions has been a point of concern \citep{1996ApJ...470.1187R}. So close to their host stars, short-period gas giants raise substantial tidal bulges on their host stars, large enough that in some cases the bulge has been detected \citep{2022A&A...657A..52B}. For host stars rotating more slowly than their short-period planetary companions revolve, the interaction between this tidal bulge and the planet transfers angular momentum from the orbit to the star, reducing the orbital distance and orbital period \citep{2008ApJ...678.1396J}. The rate at which tidal energy is dissipated within the host star determines the orbital decay rate but, for stars on the giant branch, may be comparable to the stellar luminosity \citep{2018ApJ...853L...1M}. The stellar dissipation processes, usually quantified via the tidal dissipation parameter $Q_\star$, are likely complex and remain poorly understood \citep{2014ARA&A..52..171O}, translating into orders of magnitude uncertainty in $Q_\star$. 

Once a gas giant spirals into its Roche limit, a distance determined in part by the stellar and planetary densities \citep{2013ApJ...773L..15R}, tidal disruption can occur. This disruption may proceed on a timescale set by the tidal decay rate \citep{2015ApJ...813..101V, 2016CeMDA.126..227J}, or the disruption may become unstable and proceed rapidly \citep{2003ApJ...588..509G, 2017MNRAS.465..149J}. Or, in a more dramatic case, the planet's Roche limit may lie within the star, in which case the tidally decaying planet can be directly accreted by the star \citep{2012MNRAS.425.2778M}.  

A variety of indirect observational evidence supports these theoretical expectations that short-period planets are disrupted and/or accreted by their host stars: some stars show signs of tidal- or accretion-induced spin-up \citep{2018ApJ...864...65Q}; main-sequence stars that currently host hot Jupiters tend to be younger on average than main-sequence stars that host planets less susceptible to tidal decay \citep{2019AJ....158..190H}; and some red giant stars exhibit anomalous chemical signatures that may be caused by planetary engulfment \citep{2016ApJ...829..127A}, and such signatures may also be present but short-lived for main sequence stars \citep{2023MNRAS.518.5465B}. \citet{De2023} provided the first direct detection of ongoing planetary engulfment. Based on a large-scale survey, that study reported detection of a low-luminosity optical transient lasting several days, accompanied by a months'-long infrared brightening. These signatures are consistent with engulfment of a planet between 0.1 and 10 Jupiter masses by a Sun-like star about $4\,{\rm kpc}$ from Earth.

Based on their survey detection statistics and other considerations, \citet{De2023} estimated such events occur at a rate between 0.1 and $1\,{\rm yr}^{-1}$. As discussed in \citet{2012MNRAS.425.2778M}, the engulfment rate scales with $Q_\star$: a value $Q_\star \sim 10^6$ translates into about one tidally-driven planetary accretion events per year within the Milky Way. However, the large uncertainties on $Q_\star$ mean the actual event rate is likewise highly uncertain. Moreover, $Q_\star$ likely depends on stellar structure, with later-type stars exhibiting more efficient dissipation (smaller $Q_\star$), and probably also on tidal driving frequency. 

One way to constrain $Q_\star$ and the galactic engulfment rate for exoplanet systems would be to observe tidally driven orbital decay, which would manifest as variations in transit timing. Unfortunately, tidal decay has only been definitively detected this way for one hot Jupiter, WASP-12 b \citep{2017AJ....154....4P, 2020ApJ...888L...5Y}. The period decay rate reported in \citet{2020ApJ...888L...5Y} $dP/dt = -29 \pm 2\,{\rm ms\ yr^{-1}}$ translates to $Q_\star \approx 2\times10^5$ and amounts to a change in the period of just under half a second since the planet was discovered in 2008. A recent analysis of TESS data confirmed this decay rate, reducing the error bars below $1\,{\rm ms\ yr^{-1}}$ \citep{2022AJ....163..175W}. Possible tidal decay has also been reported for several other systems, including for XO-3 b \citep{Yang_2022, 2022ApJS..259...62I}, WASP-19 b \citep{2020AJ....159..150P, 2022ApJS..259...62I}, TrES-1 b, TrES-2 b, HAT-P-19 b \citep{2022AJ....164..220H}, Kepler-1658 b \citep{2022ApJ...941L..31V}, and KELT-9 b \citep{2023A&A...669A.124H}. However, definitive confirmation will likely require additional years of observations, and knowing which planets to prioritize requires understanding how decay is detected and what parameters best suit a system to exhibit detectable decay.

Detecting tidal decay requires fitting an ephemeris to observed transit times. In the absence of tidal decay, the transit times are regularly spaced (by orbital period $P$) and increase linearly with observational epoch $E$. When there is tidal decay, transits come faster and faster over time as the orbital period declines, and an additional quadratic term proportional to $E^2$ and involving the period change $dP/dE$ appears in the ephemeris. Deciding whether a series of transit times is better modeled with a linear ephemeris with no decay or a quadratic ephemeris with decay  requires considering more than the standard reduced $\chi^2$ \citep{2002nrca.book.....P}: the quadratic ephemeris can always, in principle, result in a smaller reduced $\chi^2$ because it involves one additional model parameter than the linear ephemeris.

In recent years, astronomers have invoked the Bayesian Information Criterion BIC \citep{1978AnSta...6..461S} to judge whether a dataset supports tidal decay. This simple expression incorporates both $\chi^2$, thereby favoring models that minimize  residuals, and a term that penalizes introducing additional model parameters, thereby favoring lower-dimensional models. In this context, the BIC can be written as
\begin{equation}
    {\rm BIC} = \chi^2 + k \ln N, \label{eqn:BIC}
\end{equation}
where $N$ is the total number of data points and $k$ is the number of fit parameters, 2 for a linear fit and 3 for a quadratic fit. Generally, when comparing two models, the one with the smaller BIC is favored. For a difference in BIC between two models $\Delta {\rm BIC}$, \citet{2020ApJ...888L...5Y} pointed out that the Bayes factor, $B$, i.e., the ratio of posterior probabilities favoring the linear (no tidal decay) to the quadratic (tidal decay) model, is given by
\begin{equation}
    B = \exp\left({-\Delta {\rm BIC}/2}\right).
\end{equation}
As an example, the collection of transit timing observations for WASP-12 b considered here give $\Delta {\rm BIC} \approx 200$, favoring a model with tidal decay by a probability $\sim10^{43}$ times larger than a model without tidal decay. Given its utility, in this study, we explore the various system parameters and observational strategies that can promote detection of tidal decay, framing our analysis around the BIC. 

We focus on the effects of tidal decay on a transiting planet's ephemeris. However, other astrophysical processes can impact it as well. Orbital precession, for example, can accelerate the transit times, thereby mimicking the effects of tidal decay, at least as far as the transit is concerned \citep{2020AJ....159..150P}. Observing a planet's eclipse times can distinguish between decay and precession since the former will accelerate both transit and eclipse times, but the latter will accelerate one and decelerate the other. Both mechanisms, though, introduce curvature into the ephemeris (whether the transit or eclipse ephemeris), and the analysis presented here can be used to explore the detection of ephemeris curvature, whatever the cause (or sign). Future studies may better tailor this approach to searches for precession, line-of-sight acceleration \citep[e.g.,][]{2008A&A...480..563D}, or other ephemeris perturbations.

In what follows, we first explore what astrophysical properties for a planetary system best lend themselves to precise transit times (Section \ref{sec:Simplified_Central_Time_Uncertainties}). Then, we consider the details of fitting both linear and quadratic curves in the cases of tidal decay and no tidal decay (Sections \ref{sec:Fitting_a_Linear_Curve_to_a_Linear_Ephemeris}, \ref{sec:Fitting_a_Quadratic_Curve_to_a_Quadratic_Ephemeris}, and \ref{sec:Fitting_a_Linear_Curve_to_a_Quadratic_Ephemeris}). Finally, we apply our formulation to several hypothetical observing programs and then to real observational data for a few systems with possible tidal decay (Section \ref{sec:Applying_the_Analytic_BIC_Expression}). Throughout the analysis, we invoke the WASP-12 system as a point of comparison. Since WASP-12 is the only system with definitively detected tidal decay, the evolution over time of the various detection statistics we explore here for this system serves as a template for detecting tidal decay in other systems. 

\section{Analysis} \label{sec:Analysis}

For our analysis, we considered data for hot Jupiter and short-period brown dwarf systems from the NASA Exoplanet Archive downloaded on \DataDate\ and subject to the following requirements:
\begin{enumerate}
    \item The planet must have ``Published Confirmed'' listed in the ``Solution Type'' column.
    \item The planetary system must have listed the ratios of both the stellar radius to the semi-major axis and the planetary to stellar radius. 
    \item The orbital period $P < 3\,{\rm days}$.
    \item The planet's radius $R_{\rm p}$ lay between five times Earth's $R_{\rm Earth}$ and ten times Jupiter's $R_{\rm Jupiter}$.
    \item The planet has an estimated mass $M_{\rm p}$.
    \item The planet must have a published orbital period $P$ and transit mid-point $T_0$ (called ``Time of Conjunction'' on the Exoplanet Archive), along with corresponding uncertainties.
\end{enumerate}
The disintegrating planet WD 1856+534 b also happens to satisfy all these criteria, but we dropped it as irrelevant. In some cases, the most recent set of system parameters provided on the Exoplanet Archive did not include required values. In those cases, we used the most recent set of values that \emph{did} include everything needed. In a handful of cases, we had to calculate the orbital semi-major axes from the provided period and stellar mass. These criteria left us with \NumHJs\ systems.

\subsection{Simplified Central Time Uncertainties} \label{sec:Simplified_Central_Time_Uncertainties}
To explore the astrophysical properties that support precise transit time estimates, we start with a simplified model for the central time $t_c$ of a transit or eclipse \citep{2008ApJ...689..499C}. This model involves (among other simplifications) neglecting orbital eccentricity and limb-darkening and assuming that the transiting planet is small compared to the star and that the out-of-transit baseline is very accurately estimated. (Numerical experimentation using fully accurate transit light curves shows this simplified estimate is good to about 10\%.) \citet{2008ApJ...689..499C} defines several useful parameters related to the transit:
\begin{eqnarray}
    \tau_0 &=& \frac{R_\star}{a} \frac{P}{2\pi},\\ 
    b &=& \frac{a}{R_\star} \cos i,\\
    T &=& 2 \tau_0 \sqrt{1 - b^2},\\
    \tau &=& 2 \tau_0 \left( \frac{R_{\rm p}}{R_\star} \right) \left( 1 - b^2 \right)^{-1/2},
\end{eqnarray}
where $R_\star$ is the stellar radius, $a$ is the orbital semi-major axis, $P$ is the orbital period, $i$ is the orbital inclination, $b$ is the impact parameter, $T$ is the total transit duration (defined as the time for the planet's center to cross from limb to limb), and $\tau$ is the ingress/egress duration. We also need a transit or eclipse depth $\delta$
\begin{equation}
\delta \approx \begin{cases} 
      \left( \frac{R_{\rm p}}{R_\star} \right)^2 & {\rm transit} \\
      \left( \frac{R_{\rm p}}{R_\star} \right)^2 \frac{I_{\rm p}}{I_\star} & {\rm occultation},
   \end{cases}  
\end{equation}
where $I_{\rm p/\star}$ is the planetary/stellar disk-integrated intensity \citep{2010exop.book...55W}. 

We also need the following parameters as defined in \citealt{2008ApJ...689..499C}:
\begin{eqnarray}
    Q &=& \sqrt{\Gamma T} \frac{\delta}{\sigma},\\
    \theta &=& \frac{\tau}{T},
\end{eqnarray}
with $\Gamma$ the sampling rate for the transit observations (assumed constant) and $\sigma$ the per-point photometric uncertainty. $Q$, therefore, correlates with total signal-to-noise ratio for the transit, and $\theta$ is the ratio of the ingress/egress duration to the total transit duration. Based on these definitions, \citet{2008ApJ...689..499C} provide a simplified estimate for the uncertainty on the central time $t_{\rm c}$
\begin{equation}
    \sigma_{t_{\rm c}} = T Q^{-1} \sqrt{\theta/2}.
\end{equation}

Plugging in all the above defined parameters, we find that
\begin{equation}
   \sigma_{t_{\rm c}} = \sqrt{\frac{\tau}{2\Gamma}} \left( \frac{\sigma}{\delta} \right) = \sigma \Gamma^{-1/2} \left( \frac{R_\star}{a}\right)^{1/2} \left( \frac{P}{2\pi} \right)^{1/2} \left( \frac{R_{\rm p}}{R_\star} \right)^{-3/2} \left( 1 - b^2 \right)^{-1/4} \times 
   \begin{cases}
    1 & {\rm transit}\\
   \left( \frac{I_{\rm p}}{I_\star} \right)^{-1} & {\rm eclipse}
   \end{cases}\label{eqn:simplified_sigma_tc}
\end{equation}
Not surprisingly, the uncertainty increases with the photometric uncertainty and decreases as the transit depth and sampling rate increase. Why, though, does the mid-transit uncertainty increase with ingress/egress duration? Consider a case with a very long ingress/egress (e.g., a near-grazing transit with $b \rightarrow 1$), which corresponds to a very nearly V-shaped light curve. In that case, determining the mid-transit time relies on being able to determine when exactly the light curve goes from decreasing with time to increasing with time, with very little transition in-between. Without sufficient sampling, for example, the instant of transition would be missed, and the mid-transit time would be highly uncertain.

\begin{figure}
    \centering
    \includegraphics[width=\textwidth]{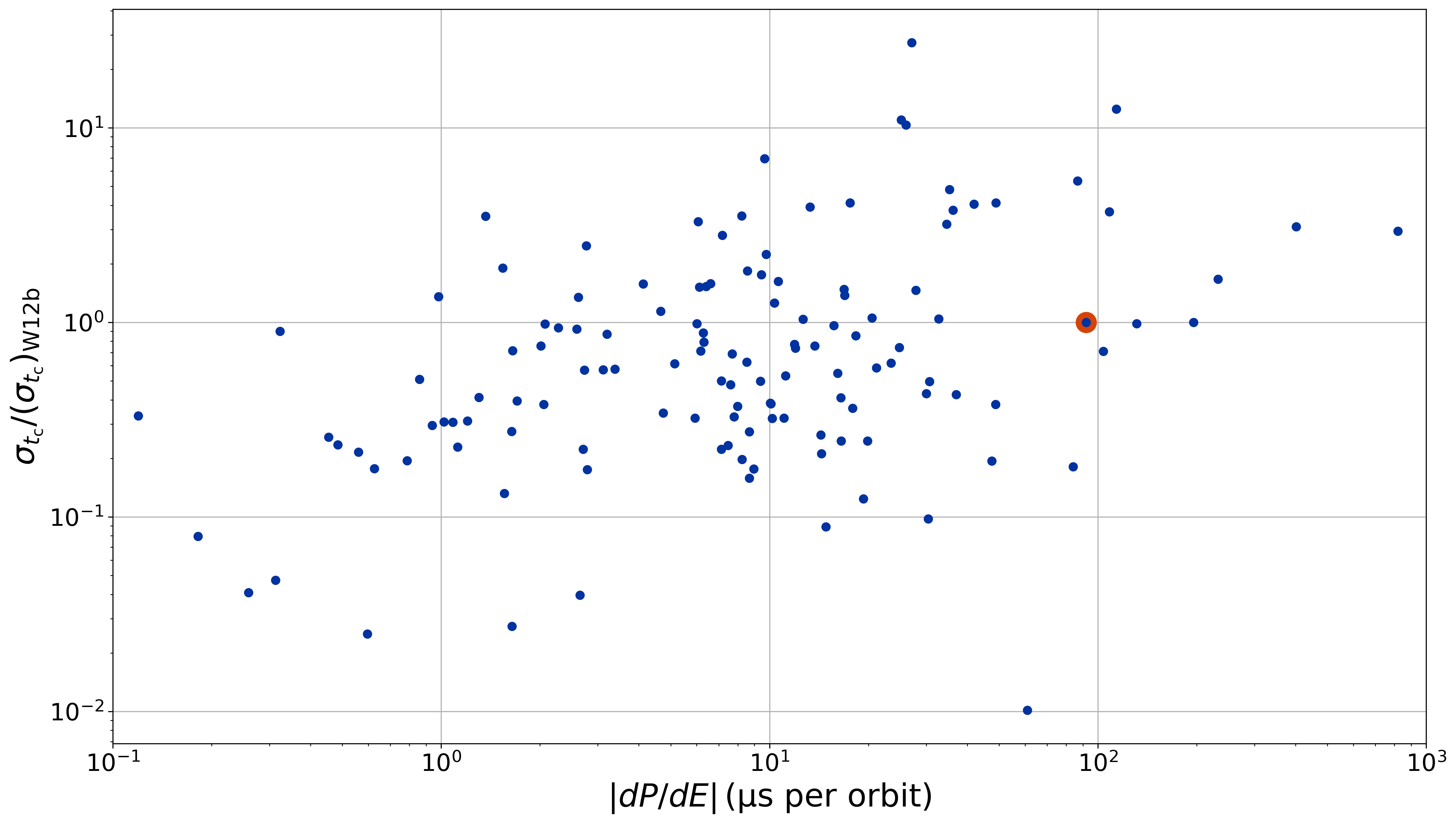}
    \caption{Simplified mid-transit time uncertainties (Equation \ref{eqn:simplified_sigma_tc}) normalized to the estimate for WASP-12 b (orange marker) vs.~the orbital decay (Equation \ref{eqn:dPdE} in Section \ref{sec:Fitting_a_Quadratic_Curve_to_a_Quadratic_Ephemeris}) for many confirmed systems.}
    \label{fig:sigma-tc_vs_dPdE}
\end{figure}

Figure \ref{fig:sigma-tc_vs_dPdE} compares estimates of transit $\sigma_{t_{\rm c}}$ for several systems to an estimate for WASP-12 b based on Equation \ref{eqn:simplified_sigma_tc}. Although it is impossible to estimate the per-point photometric uncertainty $\sigma$ for any system in general since the photometric uncertainty depends on the complex details of a particular observation, we can at least include the approximate dependence on stellar magnitude. First, we can relate the photon count rate $N$ to the stellar flux in the bandpass of observation $F$ as $N \propto F$. Then we can fold in the relationship between flux and apparent magnitude $m \propto -2.5 \log_{10} F$. Assuming Poisson statistics gives
\begin{equation}
    \sigma \propto 10^{-m/5}.\label{eqn:magnitude_correction_for_sigma}
\end{equation}

Although simplified, the calculations illustrated in Figure \ref{fig:sigma-tc_vs_dPdE} show that there are many systems that we might expect to have smaller timing uncertainties than WASP-12 b and many systems with tidal decay expected to be larger. But there are only three with both: WASP-103 b, KELT-16 b, and KELT-1 b. \citet{2022A&A...657A..52B} analyze combined ground- and space-based transit observations of WASP-103 b, realizing typical timing uncertainties about 50\% smaller than those for WASP-12 b reported in \citet{2020ApJ...888L...5Y}. However, \citet{2022A&A...657A..52B} report no detection of tidal decay but do see tidal deformation of the planet. Likewise, \citet{2023A&A...669A.124H} combine ground- and space-based data for KELT-16 b and find that the BIC favors no tidal decay but only slightly: ${\rm BIC} = 292.9$ for a constant period and ${\rm BIC} = 297.6$ for decay. Finally, \citet{2023MNRAS.521.1200B} combine 
19 transit observations for the brown dwarf system KELT-1 b and also find no evidence for tidal decay but do report possible signs of tidal synchronization of the host star's rotation. 

Having developed a sense for the range of timing uncertainties and tidal decay rates, we next turn to how transit observations are transformed into ephemerides, both those that include no decay (i.e., linear in the observational epoch $E$) and those that do include it (i.e., quadratic in $E$). 

\subsection{Fitting a Linear Curve to a Linear Ephemeris}\label{sec:Fitting_a_Linear_Curve_to_a_Linear_Ephemeris}
For a linear fit to a linear ephemeris based solely on transits, we can calculate the uncertainties on $T_0$ and $P$ using the epoch $E$ for each observed transit and the associated mid-transit time uncertainty $\sigma_{t(E)}$. We use $\sigma_{t(E)}$ to represent the actually observed uncertainty (as opposed to the analytic uncertainty for a single transit $\sigma_{t_{\rm c}}$ or the uncertainty for the predicted future transit time $\sigma_{t_{\rm tra}^{\rm pred}}$). 
The predicted time and associated uncertainty for the transit time are, respectively,
\begin{eqnarray}
    t_{\rm tra}^{\rm pred} &=& T_0 + P E,\label{eqn:t_tra_linear}\\
    \sigma_{t_{\rm tra}^{\rm pred}} &=& \sqrt{\sigma_{T_0}^2 + E^2 \sigma_{P}^2 + 2 E \sigma_{T_0, P}} \approx \sqrt{\sigma_{T_0}^2 + E^2 \sigma_{P}^2},\label{eqn:sigma_t_tra_linear}
\end{eqnarray}
Although the $\sigma_{T_0, P}$ term contributes, in practice, it is usually orders of magnitude smaller than the other terms, so we neglect it.

We can estimate the uncertainties analytically using standard linear regression \citep[cf.][]{2002nrca.book.....P}. First, define
\begin{eqnarray}
    S &=& \sum_{E \in {\rm transits}} \left( 1/\sigma^2_{t(E)}\right)\nonumber\\
    S_{E} &=& \sum_{E \in {\rm transits}} \left( E/\sigma^2_{t(E)}\right)\nonumber\\
    S_{E^2} &=& \sum_{E \in {\rm transits}} \left( E^2/\sigma^2_{t(E)}\right).
\end{eqnarray}

\begin{figure}
    \centering
    \includegraphics[width=\textwidth]{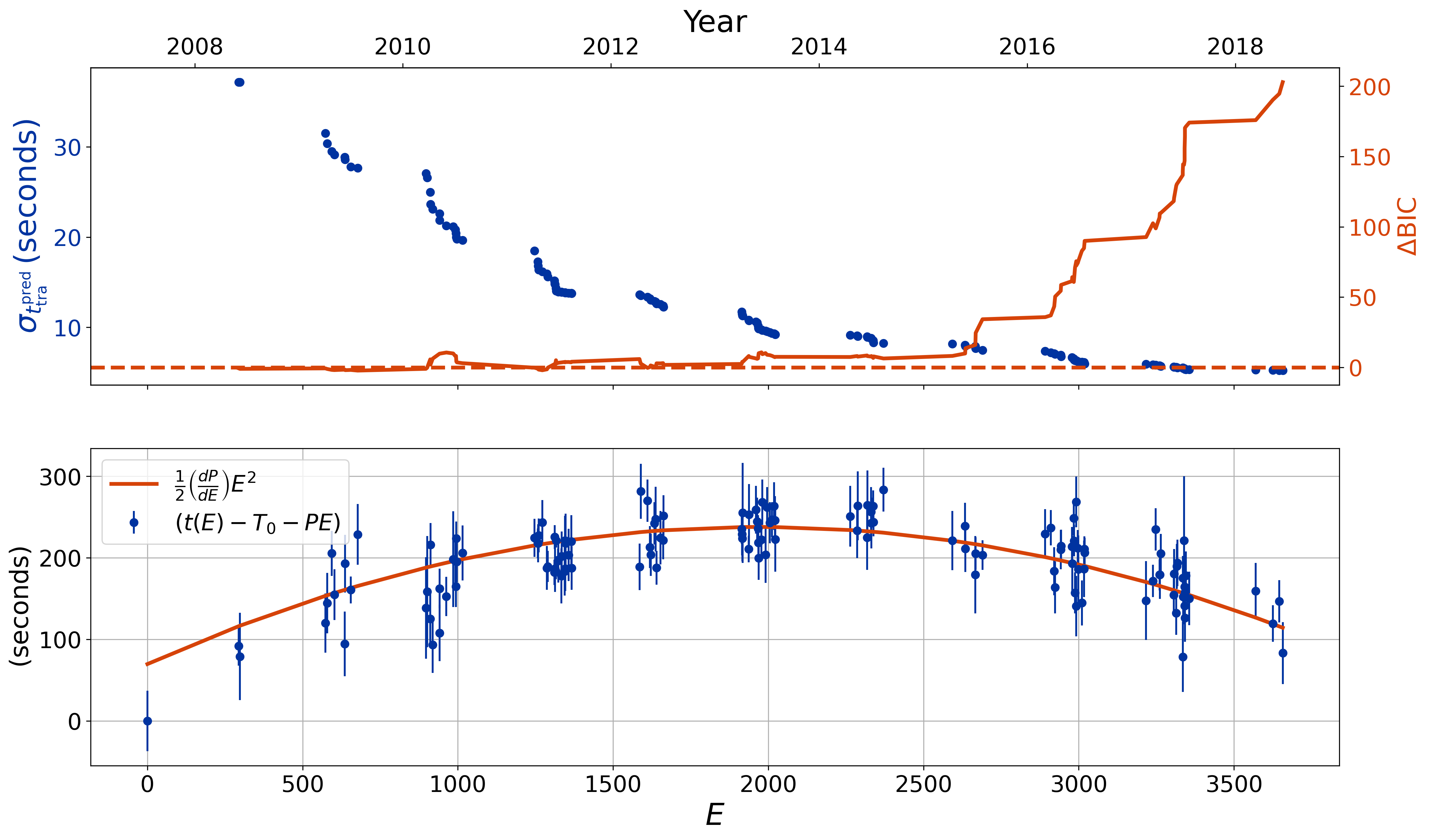}
    \caption{(top) The blue dots show the evolution of $\sigma_{t_{\rm tra}^{\rm pred}}$ for WASP-12 b during the last several years, as calculated using the data from \citet{2020ApJ...888L...5Y} and Equation \ref{eqn:sigma_t_tra_linear}. The orange curve (which uses the right y-axis) illustrates the corresponding evolution of the difference in Bayesian Information Criterion (BIC) comparing a linear ephemeris BIC(lin) to a quadratic ephemeris BIC(quad). (bottom) The blue dots show the difference between the observed transit time $t(E)$ at epoch $E$ and a linear ephemeris fit $T_0 + P E$. The orange line shows the quadratic ephemeris term using the $dP/dE$ (\YeedPdE) from \citet{2020ApJ...888L...5Y}.}
    \label{fig:Evolution of sigma_t_tra-pred for WASP-12 b}
\end{figure}

Then
\begin{eqnarray}
    \sigma_{T_0}^2 &=& \frac{S_{E^2}}{SS_{E^2} - \left( S_E \right)^2}\\
    \sigma_{P}^2 &=& \frac{S}{SS_{E^2} - \left( S_E \right)^2}.
\end{eqnarray}
Figure \ref{fig:Evolution of sigma_t_tra-pred for WASP-12 b} illustrates how adding more and more observations impacts $\sigma_{t_{\rm tra}^{\rm pred}}$ by following the history of transit observations of WASP-12 b.

Of course, if we wait for a while before observing another transit, the uncertainty for the next expected transit time will grow as in Equation \ref{eqn:sigma_t_tra_linear}. If we waited long enough $t_{\rm wait}$ that $\sigma_{t_{\rm tra}^{\rm pred}}$ grows beyond some value, then scheduling the next transit observation could be challenging:
\begin{equation}
    t_{\rm wait} = \sqrt{\sigma_{t_{\rm tra}^{\rm pred}}^2 - \sigma_{T_0}^2} \left( \frac{P}{\sigma_{P}} \right). \label{eqn:t_wait}
\end{equation}

\begin{figure}
    \centering
    \includegraphics[width=\textwidth]{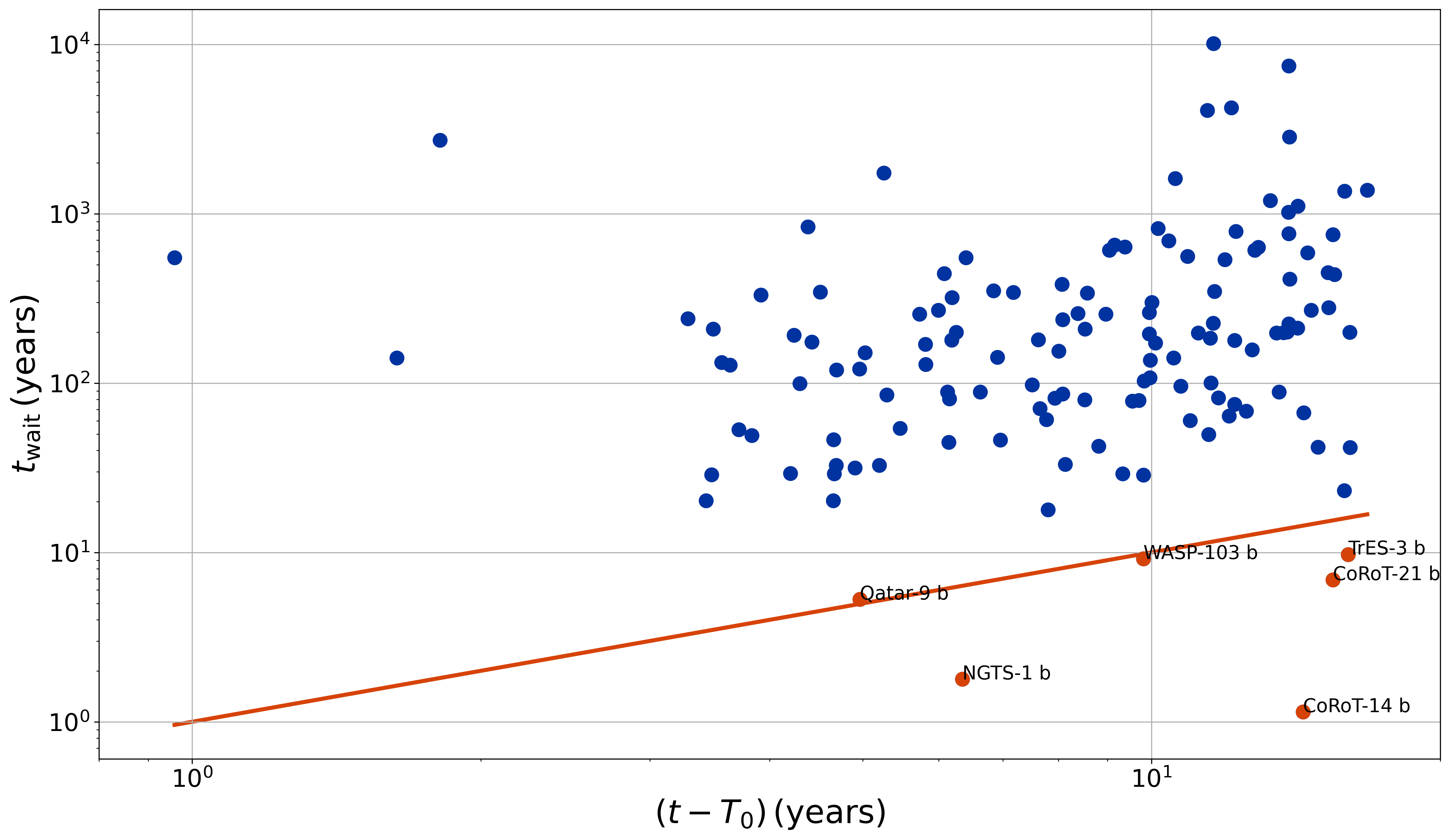}
    \caption{The time $t_{\rm wait}$ expected before the uncertainty on the linear ephemeris $\sigma_{t_{\rm tra}^{\rm pred}}$ (Equation \ref{eqn:sigma_t_tra_linear}) grows as large as the transit duration $T$ vs.~the time (as of \DataDate) since the $T_0$ value reported on the Exoplanet Archive. The orange line shows $y = x$, and the orange points are systems which lie near or below that line. For these systems, additional transit observations would likely significantly improve the linear ephemeris.}
    \label{fig:twait_vs_time-since-T0}
\end{figure}

Systems that have not been observed for $t_{\rm wait}$ are the ones for which additional transit observations would be most fruitful for improving the linear ephemeris. For the systems considered here, Figure \ref{fig:twait_vs_time-since-T0} shows the expected time we would have to wait for $\sigma_{t_{\rm tra}^{\rm pred}}$ to grow as large as the transit duration $T$ since the $T_0$ value reported on the Exoplanet Archive (as of \DataDate). Most planets have sufficiently precise linear ephemerides that we would have to wait many years before uncertainties on their expected $t_{\rm tra}^{\rm pred}$ grew as large as their transit durations, but uncertainties for a handful are likely large enough to warrant follow-up already, at least based on the Exoplanet Archive data. CoRoT-14 b, for example, has an orbital period $P = 1.51214\pm0.00013\,{\rm days}$ and $T_0 = 2454787.6702\pm0.0053$ JD \citep{2017A&A...602A.107B}, which corresponds to 2008 Nov. Over the last decade and a half, $\sigma_{t_{\rm tra}^{\rm pred}}$ has grown as large as its transit duration, $T = 1.2\,{\rm hours}$. Very near the one-to-one line, WASP-103 b was recently observed by the CHEOPS telescope, observations which actually suggest an orbital period increase rather than a decrease \citep{2022A&A...657A..52B}. However, the resulting ephemeris was too recent to have been included in our data, and so we do not consider it here. TrES-3 b is a similar case -- we did not use more recent observations \citep[e.g.,][]{2022AJ....164..198M} that would likely update its timing uncertainty and increase $t_{\rm wait}$. Determining whether individual systems require follow-up or just updated ephemerides, we leave for future work.

Finally, fitting a linear curve to a linear ephemeris would be expected to result in a BIC given by
\begin{equation}
    {\rm BIC(lin)} = \left( N - 2 \right) + 2 \ln N. \label{eqn:BIC_lin}
\end{equation}

\subsection{Fitting a Quadratic Curve to a Quadratic Ephemeris}
\label{sec:Fitting_a_Quadratic_Curve_to_a_Quadratic_Ephemeris}
For a quadratic ephemeris, the predicted time and associated uncertainty for the time of the $E$th transit are, respectively,
\begin{eqnarray}
    t_{\rm tra}^{\rm pred} &=& T_0 + P E + \frac{1}{2}\left( \frac{dP}{dE} \right) E^2,\label{eqn:t_tra_quad}\\
    \sigma_{t_{\rm tra}^{\rm pred}} &\approx& \sqrt{\sigma_{T_0}^2 + E^2 \sigma_{P}^2 + \frac{1}{4} E^4 \sigma_{dP/dE}^2},\label{eqn:sigma_t_tra_quad}
\end{eqnarray}
where we have again neglected covariance between fit parameters. 

For tidal decay involving a constant phase lag (i.e., a constant value for the star's modified tidal dissipation parameter $Q_\star$), $dP/dE$ is given by
\begin{equation}
    \frac{dP}{dE} = P \frac{dP}{dt}\approx -\left( 26\,{\rm \mu s\ per\ orbit} \right) \left( \frac{M_{\rm p}}{{\rm M_{Jup}}} \right) \left( \frac{M_\star}{{\rm M_\odot}} \right)^{-8/3} \left( \frac{R_\star}{{\rm R_\odot}} \right)^5 \left( \frac{P}{{\rm day}} \right)^{-10/3} \left( \frac{Q_\star}{10^5} \right)^{-1},\label{eqn:dPdE}
\end{equation}
where $M_{\rm p}$ is planetary mass in Jupiter masses (${\rm M_{Jup}} = 1.89813\times10^{27}\,{\rm kg}$), $M_\star$ is stellar mass in solar masses (${\rm M_\odot} = 1.989\times10^{30}\,{\rm kg}$), $R_\star$ is stellar radius in solar radii (${\rm R_\odot} = 6.957\times10^8\,{\rm m}$), and $P$ is orbital period in days. 

Figure \ref{fig:sigma-tc_vs_dPdE} compares estimates of $dP/dE$ for many systems to $dP/dE$ for WASP-12, assuming a WASP-12-like $Q_\star = 2\times10^5$. Interestingly, many systems might be expected to exhibit faster tidal decay, and many more systems likely have properties that give rise to more precise transit timing $\sigma_{t_{\rm c}}$, at least based on the simplified analytic treatment outlined in Section \ref{sec:Simplified_Central_Time_Uncertainties}. 

By analogy with the linear case, we can analytically calculate the uncertainties on the fit parameters. For this calculation, we define
\begin{eqnarray}
    S_{E^3} &=& \sum_{E \in {\rm transits}} \left( E^3/\sigma^2_{t(E)}\right),\nonumber\\
    S_{E^4} &=& \sum_{E \in {\rm transits}} \left( E^4/\sigma^2_{t(E)}\right).
    \label{eqn:SE4_definition}
\end{eqnarray}
With these definitions,
\begin{eqnarray}
    \sigma_{T_0}^2 &=& \frac{S_{E^3}^2 - S_{E^2} S_{E^4}}{\Delta},\\
    \sigma_{P}^2 &=& \frac{S_{E^2}^2 - S S_{E^4}}{\Delta},\\
    \sigma_{dP/dE}^2 &=& \frac{S_{E}^2 - S S_{E^2}}{\Delta}
    \label{eqn:quad_coeff_uncertainties}
\end{eqnarray}
where 
\begin{equation}
\Delta = -S S_{E^2} S_{E^4} + SS_{E^3}^2 + S_E^2S_{E^4} - 2S_E S_{E^2} S_{E^3} + S_{E^2}^3 \nonumber.    
\end{equation}

Fitting a quadratic curve to a quadratic ephemeris would be expected to result in 
\begin{equation}
    {\rm BIC(quad)} = \left( N - 3 \right) + 3 \ln N.
\end{equation}

\begin{figure}
    \centering
    \includegraphics[width=\textwidth]{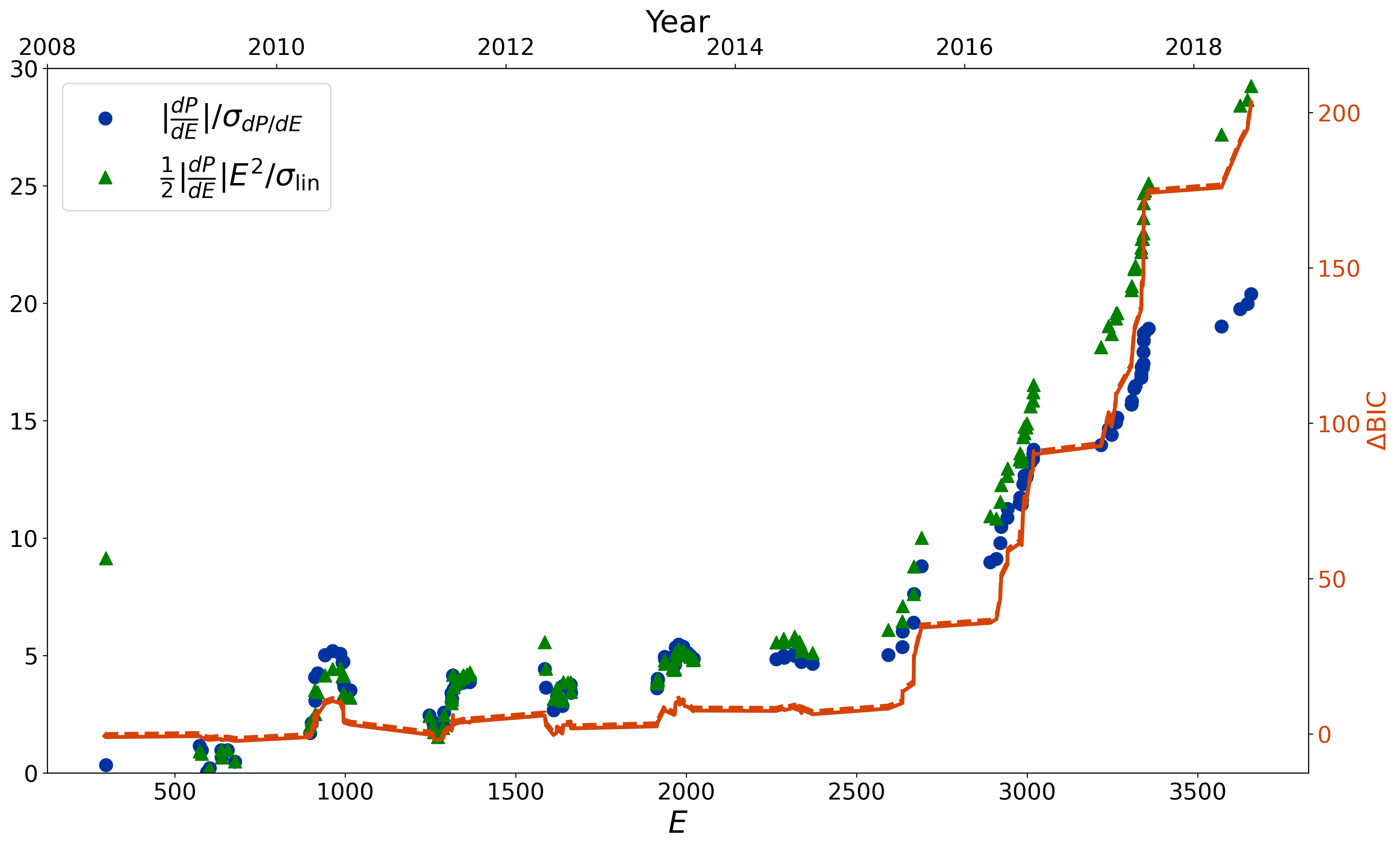}
    \caption{The blue dots show the evolution of the tidal decay signal-to-noise SNR $= |\frac{dP}{dE}|/\sigma_{dP/dE}$ for WASP-12. The green triangles show the corresponding evolution of the best-fit quadratic term $\frac{1}{2}| \frac{dP}{dE} | E^2/\sigma_{\rm lin}$. Both of these data use the left y-axis. The orange curve shows the same difference in BIC values as shown in Figure \ref{fig:Evolution of sigma_t_tra-pred for WASP-12 b} calculated numerically. These results are based on the same data as in Figure \ref{fig:Evolution of sigma_t_tra-pred for WASP-12 b} and as reported in \citet{2020ApJ...888L...5Y}. The dashed orange line shows the analytic approximation given by Equation \ref{eqn:analytic_Delta_BIC}.}
    \label{fig:quad_term_evolution_and_SNR}
\end{figure}

We can use these expressions to explore how evidence for tidal decay in the WASP-12 system mounted over the years as a template for finding other systems exhibiting tidal decay. The blue dots in Figure \ref{fig:quad_term_evolution_and_SNR} shows the evolution of the tidal decay signal-to-noise SNR for WASP-12 b alongside the comparison of the BIC for linear and quadratic fits, $\Delta {\rm BIC} \equiv {\rm BIC(lin)} - {\rm BIC(quad)}$. $\Delta {\rm BIC}$ will grow as the data favor tidal decay. Not surprisingly, as the tidal decay SNR goes up, the BIC preference for the quadratic fit increases, too. 

Another requirement for observational constraints on $dP/dE$ to be meaningful is that uncertainties on the linear portion of the ephemeris need to be small compared to the quadratic portion. Otherwise, apparent deviations from a putative linear ephemeris due to tidal decay could be attributed to the uncertainties on the linear ephemeris. This condition translates to
\begin{equation}
    \sigma_{\rm lin} \equiv \sqrt{\sigma_{T_0}^2 + E^2 \sigma_{P}^2} < \frac{1}{2} \bigg| \frac{dP}{dE} \bigg| E^2.\label{eqn:initial_condition_on_quad_term}
\end{equation}
Figure \ref{fig:quad_term_evolution_and_SNR} shows how this condition played out for WASP-12 b. The increase in $\Delta {\rm BIC}$ clearly correlates with the growth of the quadratic term in Equation \ref{eqn:initial_condition_on_quad_term}.

\begin{figure}
    \centering
    \includegraphics[width=\textwidth]{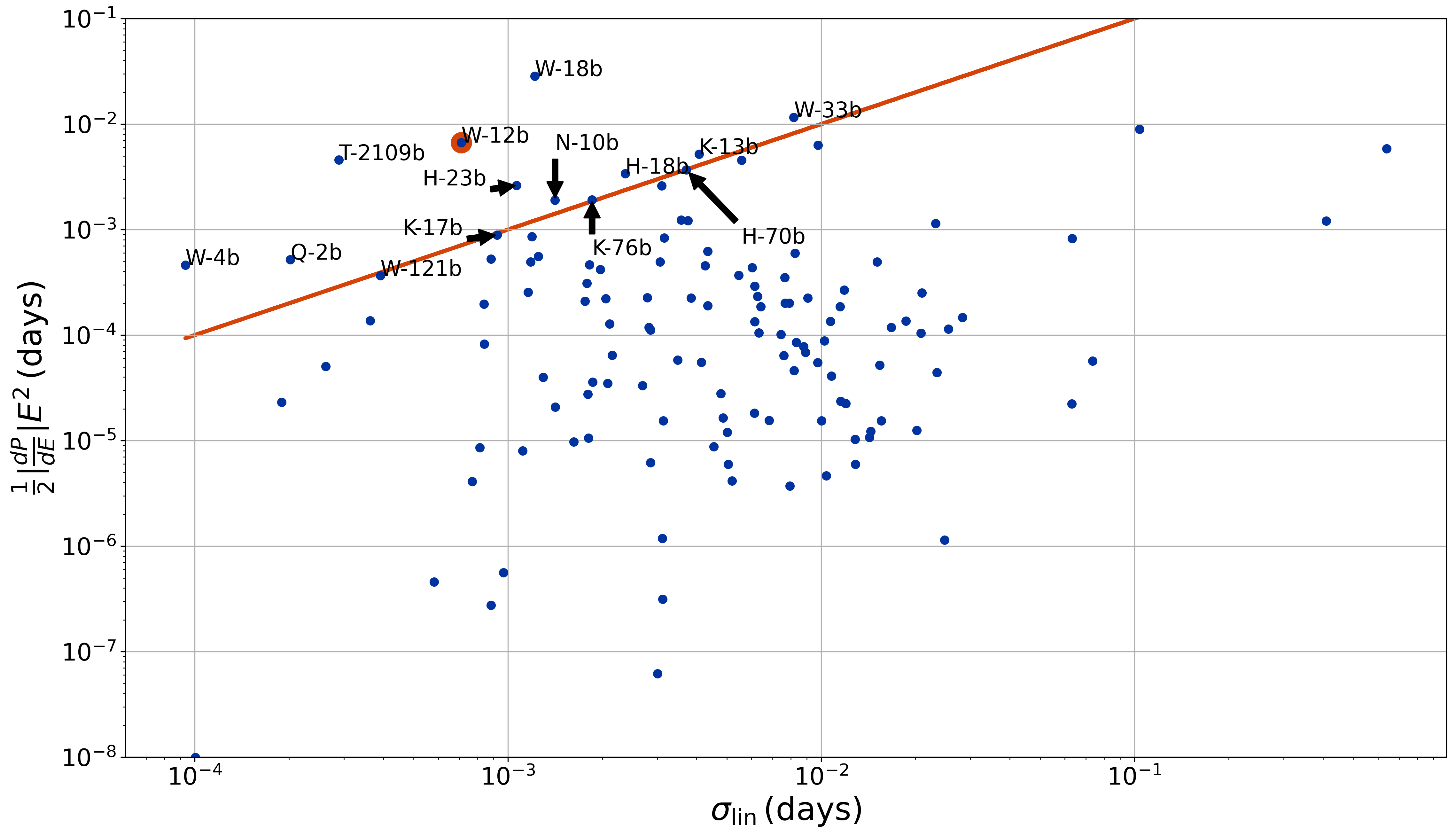}
    \caption{Cumulative change in orbital period due to tidal decay expected theoretically vs.~uncertainties on the linear ephemeris (Equation \ref{eqn:sigma_t_tra_linear}). WASP-12 b is shown with an orange circle, and the orange line shows $y = x$. Systems above that line may have experienced sufficient tidal decay since $T_0$ that it is distinguishable from uncertainties on the linear ephemeris. Individual planets near or above the line are labeled as follows: HAT-P-23 b = ``H-23b'', HATS-18 b = ``H-18b'', HATS-70 b = ``H-70b'', KOI-13 b = ``K-13b'', Kepler-17 b = ``K-17b'', Kepler-76 b = ``K-76b'', NGTS-10 b = ``N-10b'', Qatar-2 b = ``Q-2b'', TOI-2109 b = ``T-2109b'', WASP-12 b = ``W-12b'', WASP-121 b = ``W-121b'', WASP-18 b = ``W-18b'', WASP-33 b = ``W-33b'', WASP-4 b = ``W-4b''.}
    \label{fig:quad_term_vs_sigma_lin}
\end{figure}

Considering other systems, Figure \ref{fig:quad_term_vs_sigma_lin} shows the cumulative change in orbital period expected due to tidal decay for our collection of systems as compared to the uncertainty on the linear ephemeris. Systems satisfying Inequality \ref{eqn:initial_condition_on_quad_term} appear above the orange line. For example, WASP-12, the only system for which tidal decay has been definitively observed, appears above that line, along with several other systems. Several caveats should be considered in evaluating these results, including the fact that we have assumed $Q_\star = 2\times10^5$. This is the value inferred for WASP-12, which may exhibit unusually efficient tidal dissipation \citep{2017ApJ...849L..11W} and therefore may not be a representative value. For some well-observed systems, such as WASP-18b and WASP-19b, the lack of observed orbital decay to date has been used to constrain their values of $Q_\star$ to $>10^7$ and $>10^6$, respectively \citep{2022A&A...668A.114R}. Even so, the results point to several systems that merit follow-up transit observations. Several of the systems above the line in  Figure \ref{fig:quad_term_vs_sigma_lin} have been noted to exhibit period changes. HAT-P-23, for example, has $\frac{1}{2}\left(\frac{dP}{dE}\right) E^2 = -0.003\, {\rm days}$ and $\sigma_{\rm lin} = 0.002\,{ \rm days}$, while \citet{2022AJ....164..220H} reported $dP/dt = -5.2\pm5.8\,{\rm ms\ yr^{-1}}$ (which works out to $\Delta P \approx -0.004\, {\rm days}$ since $T_0$ for HAT-P-23 b).

\subsection{Fitting a Linear Curve to a Quadratic Ephemeris}\label{sec:Fitting_a_Linear_Curve_to_a_Quadratic_Ephemeris}

Finally, we consider the case of fitting a linear curve to a quadratic ephemeris. Analyzing this case is useful since it will allow us to explore how to estimate the BIC thresholds we should look for if we suspect a planet shows signs of tidal decay. (The other combination, fitting a quadratic to a linear ephemeris, would, in principle, result in a quadratic coefficient statistically consistent with zero and the same BIC expression as from Section \ref{sec:Fitting_a_Quadratic_Curve_to_a_Quadratic_Ephemeris}.) 

To start, consider the linear curve that results from fitting the quadratic ephemeris. Since $|dP/dE| \ll T_0$ and $|dP/dE| \ll P$ (where $T_0$ and $P$ are the true values for the system), we might suspect that the best-fit values, which we will call $T_0^\prime$ and $P^\prime$, would closely resemble the actual values, i.e. $T_0^\prime \approx T_0$ and $P^\prime \approx P$. Indeed, fitting a linear curve to the transit times for WASP-12 b from \citet{2020ApJ...888L...5Y} returns $T_0^\prime$ and $P^\prime$ that match $T_0$ and $P$ to better than a few parts in ten thousand. But the large collection of high quality data for WASP-12 b means that even this small disagreement is still statistically discrepant. This result comports with the results from Section \ref{sec:Fitting_a_Quadratic_Curve_to_a_Quadratic_Ephemeris}: those systems for which we have sufficient data to detect a non-zero $dP/dE$ are also those for which we have very small error bars on $T_0$ and $P$. Therefore, in order to calculate BIC for fitting a linear curve to a quadratic ephemeris, we will need also to calculate $T_0^\prime$ and $P^\prime$, which can be written as
\begin{eqnarray}
    T_0^\prime &=& T_0 + \left( \frac{dP}{dE} \right) \Delta T_0^\prime\\
    P^\prime &=& P + \left( \frac{dP}{dE} \right) \Delta P^\prime.
\end{eqnarray}
where $\Delta T_0^\prime$ and $\Delta P^\prime$ are the corrections we need to work out. As outlined in the Appendix, standard linear regression gives the following formulae for $\Delta T_0^\prime$ and $\Delta P^\prime$:
\begin{eqnarray}
\Delta T_0^\prime = \left( \frac{S_{E^2}^2 - S_{E^3} S_E}{S_{E^2}S - S_E^2} \right)\\
\Delta P^\prime = \left( \frac{S_{E^3} S - S_{E^2} S_E}{S_{E^2}S - S_E^2} \right).\label{eqn:Delta_T0_and_P_prime}
\end{eqnarray}

Now, we can calculate the resulting $\chi^2$ for fitting a linear curve to a quadratic ephemeris:
\begin{eqnarray}
\chi^2 &=& \sum_{E \in {\rm transits}} \left( \frac{t(E) - P^\prime E - T_0^\prime}{\sigma_{t(E)}} \right)^2\\\nonumber
&\approx& \frac{1}{4} \left( \frac{dP}{dE} \right)^2 \sum_{E \in {\rm transits}} \left( \frac{E^2 - \Delta P^\prime E - \Delta T_0^\prime}{\sigma_{t(E)}} \right)^2 + \left( N - 2 \right)\label{eqn:chi_squared_linear_to_quad}
\end{eqnarray}
and the difference in BIC values for a linear curve and a quadratic curve, both fit to a quadratic ephemeris as determined analytically
\begin{equation}
    \Delta {\rm BIC} = \frac{1}{4} \left( \frac{dP}{dE} \right)^2 \sum_{E \in {\rm transits}} \left( \frac{E^2 - \Delta P^\prime E - \Delta T_0^\prime}{\sigma_{t(E)}} \right)^2 - \ln N + 1.\label{eqn:analytic_Delta_BIC}
\end{equation}

Of course, given a set of already observed transits, we could easily calculate the $\Delta {\rm BIC}$. The benefit of Equation \ref{eqn:analytic_Delta_BIC} is that we can estimate the $\Delta {\rm BIC}$ expected for a sequence of \emph{planned} transit observations that have yet to be conducted (given reasonable estimates for the expected timing uncertainties). The dashed orange line in Figure \ref{fig:quad_term_evolution_and_SNR} shows how closely the analytic approximation matches the numerical result obtained by directly comparing a linear to a quadratic fit. 

\section{Applying the $\Delta {\rm BIC}$ Expression}
\label{sec:Applying_the_Analytic_BIC_Expression}
Applying Equation \ref{eqn:analytic_Delta_BIC} to the ephemerides for transiting planets can provide the likelihood of detecting tidal decay for a given planned series of transit campaigns: for an expected tidal decay rate (Equation \ref{eqn:dPdE}), when should observations be collected and how many? A comprehensive application of Equation \ref{eqn:analytic_Delta_BIC} to the suite of transiting hot Jupiters could be fruitful in these ways, but for the present paper, we confine our application to a few example cases. 

\subsection{Hypothetical Cases}

\begin{figure}
    \centering
    \includegraphics[width=\textwidth]{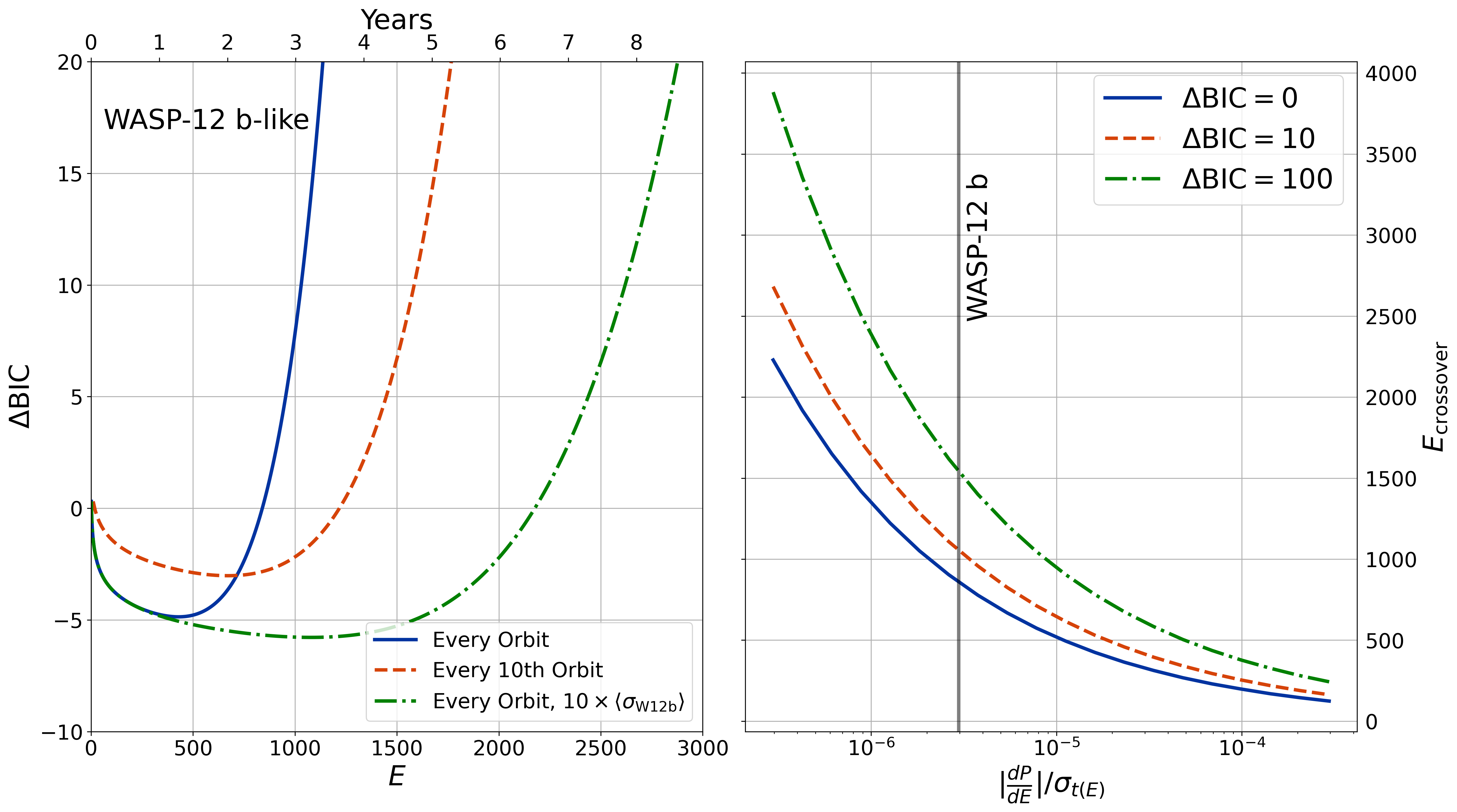}
    \caption{Application of Equation \ref{eqn:analytic_Delta_BIC} to planetary systems with WASP-12 b-like tidal decay and transit-timing uncertainties. The left panel shows the evolution of $\Delta {\rm BIC}$ assuming (1) a transit is observed every orbit (solid blue curve), (2) a transit is observed every tenth orbit (dashed orange), and (3) a transit is observed every orbit but with a timing uncertainty ten times larger than the other two cases (dash-dot green). The right panel shows the epoch $E_{\rm crossover}$ at which $\Delta {\rm BIC}$ crosses over a given value as a function of the tidal decay rate $\left( dP/dE \right)$ and transit timing uncertainty $\sigma_{t(E)}$. As a point of comparison, for WASP-12 b, $| dP/dE | / \sigma_{t(E)} \approx 3\times10^{-6}$, as shown by the vertical gray line. These curves assume optimistically that the transit for every epoch is observed up to $E_{\rm crossover}$. }
    \label{fig:Delta_BIC_predictions_hypothetical_planets_a}
\end{figure}

First, we consider hypothetical cases to gauge how effectively tidal decay could, in principle, be detected by various observing strategies for planetary systems with definite quadratic ephemerides. For many of the calculations in this section, we assumed a tidal decay rate equal to WASP-12 b's, $P/| dP/dt | = 2.3\,{\rm Myrs}$ or $dP/dE = $ \YeedPdE \citep{2020ApJ...888L...5Y} and a constant transit timing uncertainty, $\sigma_{t(E)} = {\rm const.}$ Strictly, $\Delta {\rm BIC}$ depends on transit epoch $E$ and \emph{not} on orbital period, but to give a sense for the timescales over which observational campaigns might be conducted, we assumed WASP-12 b's orbital period $P = 1.091419649\,{\rm days}$ to convert from $E$ to years.

To begin with, we consider some overly simple observational campaigns -- left panel of Figure \ref{fig:Delta_BIC_predictions_hypothetical_planets_a}. (N.B.: Throughout this section, the x- and y-axes of different panels often do \emph{not} match up.) For the blue and orange lines, that uncertainty was taken as equal to the median for the WASP-12 b dataset from \citet{2020ApJ...888L...5Y}, $\sigma_{t(E)} = \langle \sigma_{\rm W12b} \rangle = 0.00032\,{\rm days}$, while the green line shows the result for an uncertainty ten times larger ($0.0032\,{\rm days}$). The blue and green lines show how $\Delta {\rm BIC}$ would grow if we could (unrealistically) observe every transit, while the orange line shows what would happen if we observed every tenth transit. 

As previously stated, $\Delta {\rm BIC} > 0$ indicates a statistical preference for a quadratic over a linear ephemeris, and all curves in the left panel of Figure \ref{fig:Delta_BIC_predictions_hypothetical_planets_a} show $\Delta {\rm BIC}$ initially drop from zero into negative values. Intuitively, this behavior reflects the need for curvature in the ephemeris to build up over time so that the quadratic term ($\frac{1}{2} |dP/dE| E^2$ -- see Equation \ref{eqn:t_tra_quad}) grows sufficiently large that a linear regression is impacted. In other words, we have to wait for a while after a transiting planet is discovered to spot tidal decay. Equation \ref{eqn:analytic_Delta_BIC} indicates that that crossover point depends on the total number of observations $N$, the timing/frequency of those observations (the summation term), the timing uncertainty, and the tidal decay rate $dP/dE$. 

The right panel of Figure \ref{fig:Delta_BIC_predictions_hypothetical_planets_a} shows the dependence of the crossover epoch $E_{\rm crossover}$ for a desired $\Delta {\rm BIC}$ value on the ratio $|dP/dE|/\sigma_{t(E)}$, assuming every transit since a planet's discovery is observed. If, for example, $\Delta {\rm BIC} = 10$ were the goal of an observing program (dashed, orange curve) for a system with WASP-12 b-like properties ($| dP/dE | / \sigma_{t(E)} \approx 3\times10^{-6}$), then a minimum of about 1000 transits would need to be observed. Approaching the situation from the opposite direction, a survey including nearly every single transit up to $E = 1500$ and with WASP-12 b-like timing uncertainties would be expected to achieve $\Delta {\rm BIC} \approx 100$ if the system actually exhibited WASP-12 b's tidal decay. In this way, we can apply Equation \ref{eqn:analytic_Delta_BIC} to a particular observing program to determine a reasonable threshold for decay detection.

\begin{figure}
    \centering
    \includegraphics[width=\textwidth]{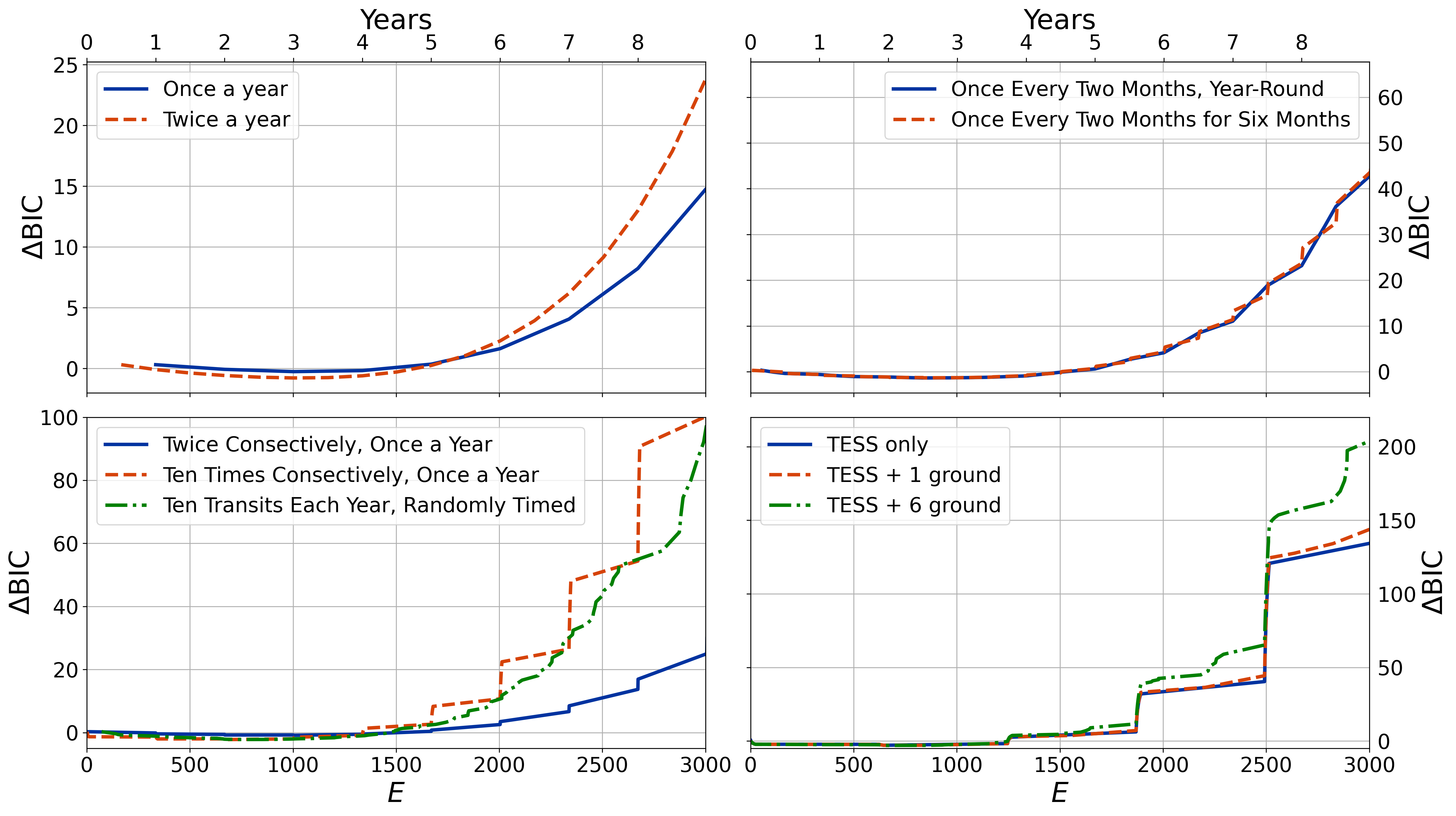}
    \caption{Application of Equation \ref{eqn:analytic_Delta_BIC} to planetary systems with WASP-12 b-like tidal decay and transit-timing uncertainties using more realistic observing strategies. The top left panel shows how $\Delta {\rm BIC}$ grows with one or two transit observations in an Earth year. The top right panel shows how $\Delta {\rm BIC}$ grows with bi-monthly transit observations during a whole (Earth) year (solid blue curve) and during an hypothetical observing season of six months with no observations during the other six months. The bottom left panel shows how $\Delta {\rm BIC}$ grows assuming two consecutive transit observations each Earth year (solid blue curve) and ten consecutive observations each Earth year (dashed orange curve). The dash-dot green curve shows the same total number of transits (ten annually) as the dashed, orange curve but randomly timed. The bottom right panel shows how $\Delta {\rm BIC}$ evolves for an observational program meant to mimic TESS observations (solid blue - ``TESS only''). That program involves about two dozen transits once every 25 TESS sectors, which span 27 days each. The dashed orange line (``TESS + ground'') shows the same observing program but with the addition of one transit observation every six months, meant to represent a ground-based observation. Observations for both programs assume the same transit timing uncertainties.}
    \label{fig:Delta_BIC_predictions_hypothetical_planets_b}
\end{figure}

Consider next somewhat more realistic observing campaigns -- Figure \ref{fig:Delta_BIC_predictions_hypothetical_planets_b}. The top left panel compares $\Delta {\rm BIC}$ growth for one transit observation and two observations in one Earth year. In about nine Earth years, the curve corresponding to twice annual observations (dashed orange) has grown to nearly twice the $\Delta {\rm BIC}$ for once annual observations (solid blue). The top right panel compares $\Delta {\rm BIC}$ growth for one observation every two months all Earth year-round and the other involving one observation every two months for six months (dashed orange). Here, the two curves weave over one another, suggesting little advantage of one program over the other. This result is not surprising since little curvature develops in the ephemeris over six months.

Next, consider the bottom left panel of Figure \ref{fig:Delta_BIC_predictions_hypothetical_planets_b}. The solid blue curve involves two consecutive transit observations each Earth year, while the dashed orange curve involves ten consecutive observations each Earth year. Not surprisingly, $\Delta {\rm BIC}$ increases much more rapidly for the latter program than for the former since, in each observing session, significantly more transits are collected. The dash-dot green curve involves the same total number of transits as the dashed, orange curve but randomly phased (i.e., not necessarily consecutive transits), illustrating that the timing of observations over a short (compared to the decay time) timescale has minimal impact on the evolution. For this particular instantiation, the final $\Delta {\rm BIC}$ ends up slightly below the dashed, orange curve, but other examples (not shown) have $\Delta {\rm BIC}$ equal to or even slightly above the dashed, orange curve, depending on exactly how the observations are timed. 

Finally, consider the bottom right panel of Figure \ref{fig:Delta_BIC_predictions_hypothetical_planets_b}. This panel shows how ground-based observations can combine with observations from a TESS-like mission to detect tidal decay. For this calculation, we first assumed every transit was observed for a WASP-12-like system during a 27-day period in each year, followed by no observations for 25 TESS sectors, and then another sequence of transits were observed, etc. The solid blue line shows this scenario. The dashed orange line shows the same program except with a single ground-based transit included every six months. The dash-dot, green line shows the same program except with six ground-based transit observations, randomly spread during a six-month observing season. Not surprisingly, $\Delta {\rm BIC}$ for the TESS + ground programs grows more quickly, demonstrating the power of combining the two approaches. The dash-dot green line (six ground-based transits) modestly significantly exceeds the solid blue line once the signal of tidal decay starts to emerge and $\Delta {\rm BIC}$ grows large, while the dotted orange line (one ground-based transit) only modestly exceeds it.

Although the approach here needs to be tailored to each specific observing program for detailed predictions, these results illustrate its general utility. They show that detecting tidal decay requires collecting regular observations and allowing sufficient time for decay to manifest. In general, a significant increase in $\Delta {\rm BIC}$ requires a significant fractional increase in the number of observations. Adding just a few more observations to an already full observing program does not make much difference unless they are judiciously timed. Two encouraging conclusions of these results: (1) an observing program that can only observe a few times a year can still have an impact, and (2) it may be more worthwhile to double the number of candidates, focusing on the planets most likely to exhibit decay, than to double the number of transits observed for a given planet if a program already involves several observations in a year.

\subsection{Real Cases}
\begin{figure}
    \centering
    \includegraphics[width=\textwidth]{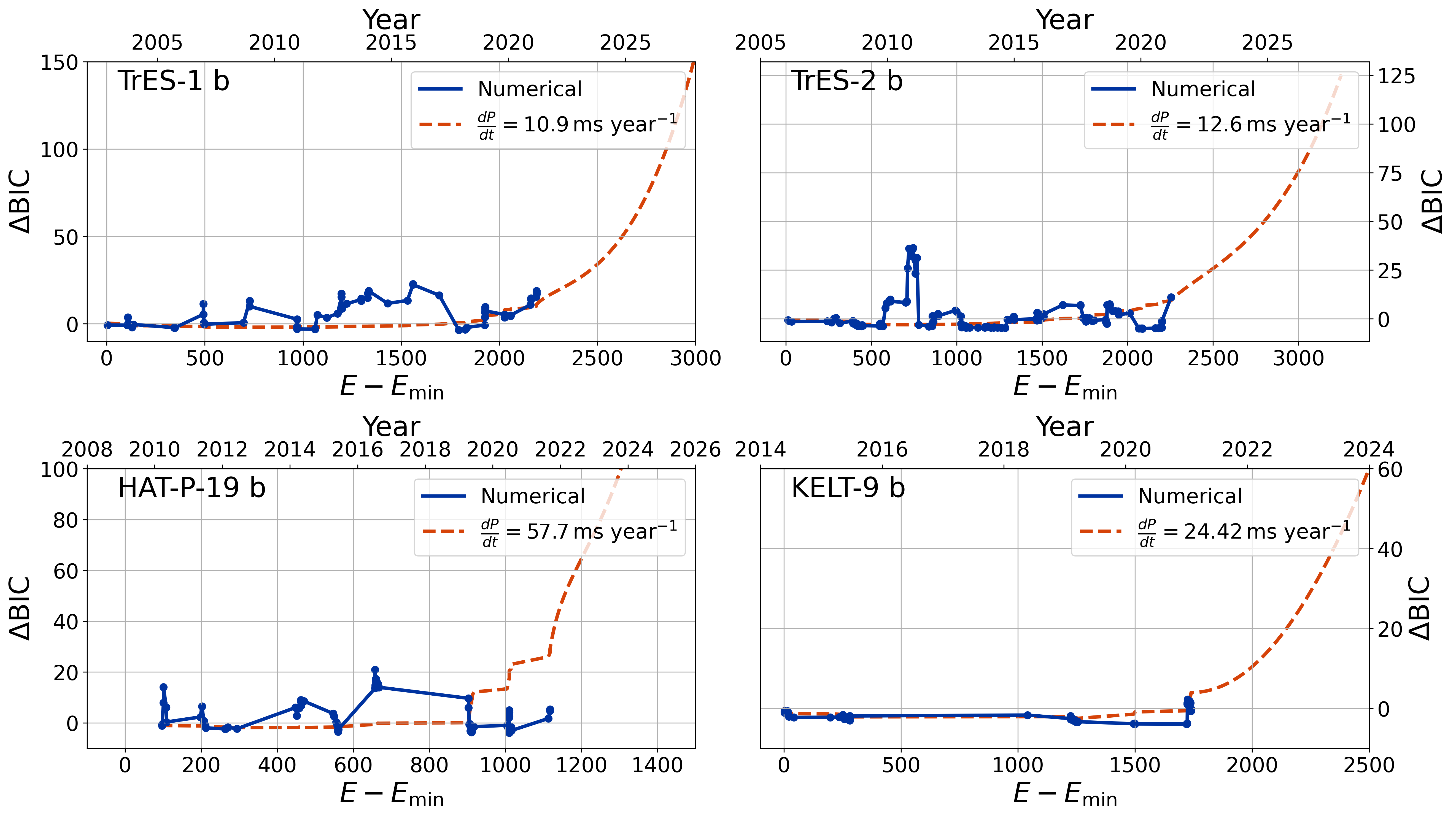}
    \caption{Evolution of $\Delta {\rm BIC}$ for several real planets. The solid blue ``Numerical'' lines show the evolution based on previously published observations, while the dashed orange lines show the evolution based on Equation \ref{eqn:analytic_Delta_BIC}. For points in that latter calculation beyond the previously published observations, we assumed $\sigma_{t(E)}$ equal to the average uncertainty for the previously published observations. We also assumed the transits were observed at a cadence equal to the median cadence for the previously published observations. For example, previous transit observations of TrES-2 b have been conducted typically once every four orbits. Observations for TrES-1 b, TrES-2 b, and HAT-P-19 b come from \citet{2022AJ....164..220H}, and observations for KELT-9 b come from \citet{2023A&A...669A.124H}. For this plot, we have subtracted the minimum reported epoch $E_{\rm min}$ from $E$. }
    \label{fig:Real_Planets}
\end{figure}

Finally, we consider real systems -- Figure \ref{fig:Real_Planets}. These examples all involve systems for which possible tidal decay has been reported. We take observations for TrES-1 b, TrES-2 b, and HAT-P-19 b from \citet{2022AJ....164..220H} and observations for KELT-9 b from \citet{2023A&A...669A.124H}. For the calculations in this section, we take the correct (and variable) transit timing uncertainties and the corresponding orbital periods to convert epoch $E$ to (Earth) years. To extrapolate the $\Delta {\rm BIC}$ evolution forward in time, we assume that observations continue with the same median frequency as before. For example, TrES-2 b has been observed every four orbits, and so we assume that same observing cadence going on past 2020.

\citet{2022AJ....164..220H} analyzed transit times reported on the Exoplanet Transit Database\footnote{\url{http://var2.astro.cz/ETD/}} and, for TrES-1 b, a tidal decay rate $dP/dt = -10.9 \pm 2.1\, {\rm ms\ yr^{-1}}$ was favored over a constant period by $\Delta {\rm BIC} = 9.7$. The left panel of Figure \ref{fig:Real_Planets} shows a good match between the numerical and analytic estimates for $\Delta {\rm BIC}$, and, assuming the nominal tidal decay rate, the analytic estimate suggests $\Delta {\rm BIC}$ ought to exceed 50 within the next few years. However, it may take until about 2030 to reach the same level as reported for WASP-12 b in \citet{2020ApJ...888L...5Y} -- not surprising, given that WASP-12 b's $dP/dt$ is about twice as large.

Moving next to TrES-2 b, \citet{2022AJ....164..220H} estimated $dP/dt = -12.6 \pm 2.4\, {\rm ms\ yr^{-1}}$ with tidal decay favored at $\Delta {\rm BIC} = 8.3$. Again, Figure \ref{fig:Real_Planets} shows a good match between the numerical and analytic estimates (albeit with considerable scatter in the ``Numerical'' estimate). Again, the smaller $dP/dt$ than WASP-12 b's means $\Delta {\rm BIC}$ grows more slowly and may not exceed 50 until 2025. Like the TrES-1 data, the TrES-2 observational data show significant statistical fluctuations in $\Delta {\rm BIC}$ -- between $E - E_{\rm min}= 500$ and 1000, $\Delta {\rm BIC}$ climbed rapidly before settling back toward zero. The dashed, orange line shows that such a rapid increase would not have been expected so soon after the planet's discovery. It remains to be seen whether the recent upward tick in $\Delta {\rm BIC}$ seen in the most recent data represents the beginning of true increase or whether it too is another statistical fluctuation, although the upward tick is consistent with expectations.

For HAT-P-19 b, \citet{2022AJ....164..220H} reported $dP/dt = -55.2 \pm 7.2\, {\rm ms\ yr^{-1}}$ and $dP/dE = 606\,{\rm \mu s\ orbit^{-1}}$, with tidal decay favored at $\Delta {\rm BIC} = 8.3$. This $dP/dE$ value is almost seven times that for WASP-12 b, and the analytic $\Delta {\rm BIC}$ reflects this, with a value predicted to exceed that for WASP-12 b by 2024. However, the ``Numerical'' estimate shows considerable non-monotonicity, reversing direction and sign several times during the observational baseline. The ``Numerical'' $\Delta {\rm BIC}$ appears to significantly under-perform the analytic estimate. In this context, the right panel of Figure \ref{fig:Delta_BIC_predictions_hypothetical_planets_a} is particularly useful. The average timing uncertainty for the HAT-P-19 b data is $0.0007296
\,{\rm days}$, about twice the average for WASP-12 b ($0.00032\,{\rm days}$). If all orbits since discovery had been observed, by $E - E_{\rm min} = 1000$, we would expect $\Delta {\rm BIC}$ to exceed 100 (dash-dot green line in the right panel of Figure \ref{fig:Delta_BIC_predictions_hypothetical_planets_a}). Of course, not every orbit of HAT-P-19 b \emph{has} been observed, but the fact that the ``Numerical'' $\Delta {\rm BIC}$ does not yet exceed 10 suggests that perhaps the tidal decay reported for HAT-P-19 b is spurious.

Finally, \citet{2023A&A...669A.124H} combined ground-based, Spitzer, TESS, and CHEOPS transits and eclipses of the ultra-hot Jupiter KELT-9 b, which orbits a star at the A/B stellar type boundary, and reported a possible decay rate $dP/dt = -24.42 \pm 10.66\,{\rm ms\ yr{-1}}$ with $\Delta {\rm BIC} = 8.4$ (although the data show a preference, $\Delta {\rm BIC} = 13.2$, for apsidal precession). Figure \ref{fig:Real_Planets} shows the ``Numerical'' $\Delta {\rm BIC}$ only became positive with the most recent observations before nosing back to 0 with the very last observation. The analytic curve suggests $\Delta {\rm BIC}$ would not have been expected to cross zero until recently anyway and that it might surpass 20 in 2023. \citet{2022ApJS..259...62I} also considered TESS observations of KELT-9 b and found no evidence for decay. Continued monitoring, especially observations of planetary eclipses, seems likely to resolve whether the system actually experiences tidal decay, which would be especially surprising since A/B stars are not expected to exhibit significant tidal dissipation \citep{2014ARA&A..52..171O}.

What to make of all these comparisons? One key conclusion is that statistical fluctuations in $\Delta {\rm BIC}$ often appear and may falsely hint at tidal decay. Continued, sustained growth in $\Delta {\rm BIC}$ is probably required to confidently report detection of tidal decay. A calculation like that depicted in the right panel of Figure \ref{fig:Delta_BIC_predictions_hypothetical_planets_a} tailored for a specific campaign provides a way of assessing the threshold $\Delta {\rm BIC}$ beyond which tidal decay may be plausible.

\section{Discussion and Conclusions}
\label{sec:Discussion_and_Conclusions}
The approach presented here allows observers to plan observational programs to maximize the possibility for detecting tidal decay while minimizing the required resources. This approach is framed in such a way that it does not, in principle, even require observations to make useful predictions: if observers have estimates for the expected transit timing uncertainty (Equation \ref{eqn:simplified_sigma_tc}) and tidal decay rate (Equation \ref{eqn:dPdE}), along with a planned observational sequence (which orbital epochs will be observed), Equation \ref{eqn:analytic_Delta_BIC} provides a way of estimating the likelihood for detecting tidal decay. 

Naturally, this approach comes with important caveats and limitations. For instance, we have assumed a linear regression approach, but real transit data may have complex and asymmetric uncertainties \citep{2022AJ....164..220H} for which such an approach is only an approximation. Our approach also assumes the BIC provides an accurate means for comparing models with tidal decay and those without. However, the BIC is only valid for sample sizes much larger than the number of model parameters \citep{1978AnSta...6..461S}. Fortunately, seeking signs of tidal decay necessitates a large number of observations, and so this requirement is likely always fulfilled in this context. Finally, we have limited our scope to tidal decay and have not explicitly considered other astrophysical processes that can affect the ephemeris. But our approach should apply to any mechanism that introduces a quadratic term into the ephemeris, so it should be able to capture the impact of precession on both transit and eclipse timings \citep{2010exop.book...55W}. Our method can likely be extended to consider other simple ephemeris effects, as long as they can be readily captured by a linear regression approach. 

Another potentially fruitful extension would be to incorporate a more sophisticated relationship between stellar properties and tidal decay rate. Studies of stellar tides suggest that the deeper convective zones in later-type (i.e., cooler) stars tend to promote tidal dissipation, resulting in smaller tidal dissipation parameters $Q_\star$ \citep[e.g.,][]{2022ApJ...927L..36B}, which would tend to recommend their planetary systems as good targets for detecting tidal decay. Main-sequence, cooler stars tend to be smaller, too, giving deeper transits and therefore smaller timing uncertainties (Equation \ref{eqn:simplified_sigma_tc}). On the other hand, main-sequence cooler stars are dimmer, which tends to inflate the photometric uncertainty (Equation \ref{eqn:magnitude_correction_for_sigma}). Figuring out how to thread this needle and choose the best set of stars to optimize decay detection should be the subject of future work.

As discussed in Section \ref{sec:Introduction}, detecting tidal decay is critical for constraining $Q_\star$ and planetary engulfment rates. As pointed out in \citet{2012MNRAS.425.2778M}, the engulfment rate should scale roughly as $Q_\star$, and if many stars had $Q_\star$ values as small as WASP-12's, we might expect a galactic engulfment rate as large as $18\,{\rm yr}^{-1}$. This value is likely overly large since there is reason to believe WASP-12 has unusually dissipative tides \citep{2019MNRAS.482.1872B}. Determining the actual rate is important for future surveys for engulfment signatures since, the larger the rate, the fewer stars would need to be monitored to catch engulfment. By efficiently directing searches for tidal decay, the approach outlined here would feed forward to guide surveys to search for engulfment as well. 

\begin{acknowledgments}
We thank an anonymous referee for thoughtful feedback. This study was supported by a grant from NASA's Exoplanet Research Program.
\end{acknowledgments}

\appendix
\label{sec:appendix}
We start with an ephemeris that resembles Equation \ref{eqn:t_tra_quad}:
\begin{equation}
    t(E) = T_0 + P E + \frac{1}{2} \left( \frac{dP}{dE} \right) E^2 + \Phi(E),
\end{equation}
where $\Phi(E)$ represents the scatter associated with point $E$. We assume $\langle \Phi(E) \rangle \approx 0$. Applying standard linear regression \citep[cf.][]{2002nrca.book.....P}, we have
\begin{eqnarray}
    T_0^\prime &=& \frac{S_{E^2} S_t - S_E S_{Et}}{SS_{E^2} - S_E^2},\\\nonumber
    P^\prime &=& \frac{S S_{Et} - S_E S_t}{SS_{E^2} - S_E^2}\label{eqn:T0_and_P_prime}
\end{eqnarray}
where 
\begin{eqnarray}
S_t &=& \sum_{E \in {\rm transits}} \left( t(E)/\sigma_{t(E)}^2 \right) \\\nonumber
& \approx & \frac{1}{2} \left( \frac{dP}{dE} \right) S_{E^2} + P S_E + T_0 S \\\nonumber
S_{Et} &=& \sum_{E \in {\rm transits}} \left( t(E) E/\sigma_{t(E)}^2 \right) \\\nonumber 
& \approx & \frac{1}{2} \left( \frac{dP}{dE} \right) S_{E^3} + P S_{E} + T_0 S_E.
\end{eqnarray}
We can then incorporate these expressions into Equation \ref{eqn:T0_and_P_prime} and separate out the terms involving $dP/dE$ to arrive at Equation \ref{eqn:Delta_T0_and_P_prime}.

\software{astropy \citep{2013A&A...558A..33A,2018AJ....156..123A}, matplotlib \citep{Hunter:2007}, numpy \citep{harris2020array}, scipy \citep{2020SciPy-NMeth}}  

\bibliography{sample631}{}
\bibliographystyle{aasjournal}

\end{document}